\documentstyle[amstex,prd,aps,graphicx,epsf,tighten]{revtex}

\renewcommand{\baselinestretch}{2}

\newcommand{\bra}{\left\langle}
\newcommand{\ket}{\right\rangle}

\newcommand{\mpc}{\mbox{$h^{-1}$~Mpc}}

\begin{document}
\title{Averaging and Cosmological Observations}
\author{A.A. Coley\dag}
\address{\dag\ Department of Mathematics and Statistics,\\
Dalhousie University, Halifax, Nova Scotia}

\maketitle

\begin{abstract}

The gravitational field equations on cosmological scales are
obtained by averaging the Einstein field equations of general
relativity. By assuming spatial homogeneity and isotropy on the
largest scales, the local inhomogeneities affect the dynamics
though correction (backreaction) terms, which can lead to
behaviour qualitatively and quantitatively different from the
Friedmann-Lema\^{i}tre-Robertson-Walker models. The effects of
averaging on cosmological observations are discussed. It is argued
that, based on estimates from observational data, the backreaction
(and, in particular, the averaged spatial curvature) can have a
very significant dynamical effect on the evolution of the Universe
and must be taken into account in observational cosmology.

\end{abstract}

\newpage

The averaging problem in cosmology is perhaps the most important
unsolved problem in mathematical cosmology. It has considerable
importance for the correct interpretation of cosmological data.
Cosmological observations
\cite{Riess:2006fw,Spergel:2006hy,bennett,sdss06}, based on the
assumption of a spatially homogeneous and isotropic
Friedmann-Lema\^{i}tre-Robertson-Walker (FLRW) model plus small
perturbations are usually interpreted as implying that there
exists dark energy, the spatial geometry is flat, and that there
is currently an accelerated expansion giving rise to the so-called
$\Lambda$CDM-concordance model with $\Omega_m \sim 1/3$ and
$\Omega_{de} \sim 2/3$. Although the concordance model is quite
remarkable, it does not convincingly fit all data
\cite{bennett,sdss06,BAO,Sarkar}. Unfortunately, if the underlying
cosmological model is not a perturbation of an exact flat FLRW
solution, the conventional data analysis and their interpretation
is not necessarily valid.

The correct governing equations on cosmological scales are
obtained by averaging the Einstein equations of general
relativity. By assuming spatial homogeneity and isotropy on the
largest scales, the inhomogeneities affect the dynamics though
correction (backreaction) terms, which can lead to behaviour
qualitatively and quantitatively different from the FLRW models.
There are a number of theoretical approaches to the averaging
problem. In the approach of Buchert \cite{buch} a cosmological
space-time splitting is employed and only scalar quantities are
averaged, whereby the  Hamiltonian constraint and the Raychaudhuri
equation can be replaced by their spatially averaged counterparts
plus an integrability condition \cite{buch}. The Buchert approach
is heuristic in that it is necessary to specify one more condition
(not obtained from the field equations) in order to obtain a
closed system of equations. The averaged Hamilton constraint can
be written:
\begin{equation}
\label{hamilton1} \Omega_m + \Omega_{\cal R} + \Omega_{\cal Q} +
\Omega_{de}     \;=\;1\;\;,\label{cosmic}
\end{equation}
where $\Omega_m = \frac{8\pi G \bra {\rho} \ket }{3 H^2}$,
$\Omega_{\cal R} = -\frac{{\cal R}}{3 H^2}$, $\Omega_{\cal Q} =
\frac{{\cal Q}}{3 H^2}$, and $\Omega_{de}$ is the
Hubble-normalized dark energy contribution (e.g.,
$\Omega_{\Lambda} \sim \frac{\Lambda}{6 H^2}$, where $\Lambda$ is
the cosmological constant), and $H(t)={\dot a_D}/a_D$ (where $a_D$
is the averaged scale factor in some domain $D$). We define
$\Omega_{tot} \equiv \Omega_m + \Omega_{de}$, so that
$|\Omega_{tot}-1|$ is given in terms of ${\cal Q}$, the kinematic
variance (backreaction) term and ${\cal R}$, the averaged spatial
curvature term (which is not necessarily isotropic).

The macroscopic gravity (MG) approach is an exact approach which
gives a prescription for the correlation functions that emerge in
an averaging of the Einstein field equations \cite{Zala}. In
\cite{CPZ} the MG equations were explicitly solved in a FLRW
background geometry and it was found that the correlation tensor
is of the form of a spatial curvature. This result was confirmed
in subsequent work in which  the spherically symmetric Einstein
equations were explicitly averaged \cite{CPpaper}, and is
consistent with the work of Buchert \cite{buch} (in the Newtonian
limit and for exact spherically symmetric spaces
\cite{Buchert:1999pq}), with the results of averaging an exact
Lema\^{\i}tre-Tolman-Bondi (LTB) spherically symmetric dust model
\cite{CPpaper}, and with results from linear perturbation theory
\cite{Bild-Futa:1991}.

There is no question that the backreaction effect is real
\cite{NewRas,NewLiS,buch,CPZ,CPpaper}. Spatial curvature must be
taken into account in observational cosmology. The only question
remaining is the potential significance of the resulting effect.
However, even a small backreaction would be of importance; for
example, a non-zero curvature, even at the 1 \% level (i.e.,
$|\Omega_{\cal R }|\sim .01$), would have a significant effect on
observations (for redshifts $z \geq 0.9$) \cite{dunkley}. This has
been verified in \cite{IT06}, where values of $|\Omega_{\cal R }|
\sim 0.05$ or larger were found to be consistent with observation
(if one allows for a varying dark energy equation of state
parameter).

The Wilkinson Microwave Anisotropy Probe (WMAP) has  reported
$\Omega_{tot} = 1.02 \pm 0.02$ \cite{bennett}. In $\Lambda$CDM
models type Ia supernovae data alone \cite{Riess:2006fw} prefers a
slightly closed Universe \cite{alam}. Taken at face value this
suggests $\Omega_{\cal R} = 0.02$. Models with non-negligible
spatial curvature have been the subject of renewed interest
recently \cite{IT06}. There are other possible (non-dark energy)
explanations for the SNIa data; indeed, it has been argued that
models with no dark energy are consistent with supernova data and
WMAP data (e.g., see \cite{Sarkar} and references within).
Combining these observational data with Large Scale Structure
(LSS) observations such as the Baryon Acoustic Oscillations (BAO)
data can put stringent limits on the curvature parameter in the
context of adiabatic $\Lambda$CDM models \cite{BAO}. However,
these data analyses are very model- and prior-dependent
\cite{Shapiro:2005}, and the assumptions may be unjustified (or
even inconsistent) and care is needed in the interpretation of
data.

Using both the Buchert equations and cosmological perturbation
theory \cite{Rasanen:2003fy}, it has been found  that $\cal Q$ is
a pure second order term and $\cal R$ is of first order \cite
{NewLiS}. The exact results of\cite{CPZ,CPpaper} are consequently
consistent with the results of perturbation theory and the results
of Buchert \cite{NewLiS,NewRas} in the following sense. If ${\cal
Q}$ is second order and  ${\cal R}$ is first order, then we can
solve the integrability condition to each order of approximation
separately. To first order, if ${\cal Q} =0$, then ${\cal
R}\propto k_D/a_D{^2}$, and the averaged spatial curvature evolves
like a constant--curvature model. Consequently, the exact results
apply at first order. ${\cal R}$ is given in terms of the
integration constant $k_D$, which depends on the averaging scale,
the scale of inhomogeneities and the matter distribution density
contrast. The dynamical effects are determined through
$\Omega_{\cal R}$.

At second order we obtain effects that can be modelled by
Buchert's heuristic approach, which can be thought of as arising
from violations of the approximations and assumptions used in the
exact approach. If ${\cal Q}$ is non-zero, in addition to the
possible second order effects, ${\cal R}$ evolves differently to
$a_D^{-2}$ and this can, in turn, significantly affect
cosmological observations \cite{buch}. Indeed, since the kinematic
backreaction $\Omega_{\cal Q}$ decays more slowly than
$\Omega_{\cal R}$ (and $\Omega_m$), in the course of expansion the
kinematic backreaction becomes dynamically more important and can
become significant when the structure formation process ``injects
backreaction'' (which could then be a possible explanation of the
coincidence problem) \cite{NewRas}. We may expect the largest
deviations from the exact approach to arise from the fact that the
Universe is not FLRW on a particular scale of observations. If
$L_s$ is the scale of averaging and $L_{F}$ is an appropriate
scale on which the Universe is FLRW, then ${\cal Q} \rightarrow 0$
as $L_s/L_{F}\rightarrow 1$. For small $L_s/L_{F}$, ${\cal Q} \sim
(L_s/L_{F})^2 {\cal R}$. A typical scale $L_s\sim$ 50 \mpc, a
reasonable homogeneity scale is $L_h \gtrsim$ 200 \mpc
\cite{Hogg:2004} and the Hubble radius $L_H\sim$ 3000 \mpc. For
these values, $(L_s/L_{F})^2$ is typically in the range $0.01$ to
$0.1$.

Let us consider the size of this affect in more detail. Robust
estimators for intrinsic curvature fluctuations using
realistically modelled clusters and voids in a Swiss-cheese model
indicates that the dark energy effects can be reduced by up to
about 30 \% \cite{buch,hellaby:volumematching}. Hence the regional
Friedmann curvature estimated on the regional Hubble scale can be
large. A rough order of magnitude estimate for the variance
implied by the observed density distribution of voids implies
$|\Omega_{\cal Q}|\lesssim 0.2$ \cite{NewRas}. Since the value of
the backreaction term ${\cal Q}$ depends on the velocity field
inside the domain $D$, it has been suggested that
peculiar--velocity catalogues may offer an alternative way of
estimating ${\cal Q}$ \cite{Buchert:1999pq}. The backreaction
parameter has also been estimated in the framework of Newtonian
cosmology \cite{Buchert:1999pq}; it was found that the
backreaction term can be quantitatively small in sufficiently
large expanding domains of the Universe (e.g., $\Omega_{\cal
Q}=0.01$), but the dynamical influence of a non--vanishing
backreaction on the other cosmological parameters can be large.
Therefore, $\Omega_{\cal R} + \Omega_{\cal Q}$ can, in principal,
be quite large.

This has suggested a possible scenario in which $\Omega_{\cal R} +
\Omega_{\cal Q} \sim 0.2-0.5$. In this case it is possible for the
observed acceleration to be explained (by backreaction) using the
standard interpretation of observational data without resorting to
dark energy. However, it is probably fair to say that this
possible scenario is not supported by most researchers in the
field \cite{buch,NewRas}.

There is a heuristic argument that $\Omega_{\cal R}\sim 10^{-2}$,
which is consistent with CMB observations. Let us assume that
$\Omega_{\cal R} \sim \epsilon$. We can estimate ${\cal Q} \sim
\bra {\sigma^2}\ket_D$, where ${\sigma_D}$ is the fluctuation
amplitude; then $\Omega_{\cal Q}\sim \bra {\sigma^2}\ket_D/H^2_D
\sim \bra {\delta}\ket_D \sim \epsilon^2$ (where $\delta$ is the
density contrast) \cite{NewLiS}. From CMB data we then have that
$|\Omega_{\cal Q}|\sim 10^{-5}$ \cite{EGS}. Consequently, we have
that $|\Omega_{\cal R}| \lesssim 10^{-2}$. Indeed, from
perturbation theory the dominant local corrections to redshift or
luminosity distance go as $\nabla \Phi$, which is only suppressed
as ($L/L_H$) (rather than go as the Newtonian potential $\Phi$ as
naively expected, which is suppressed as $(L/L_H)^2)$
\cite{Rasanen:2003fy}; for scales as large as a few hundred Mpc,
the suppression is therefore mild, with $\Omega_{\cal Q} \sim
10^{-5}$ \cite{NewRas}.

This suggests a second possible scenario in which $|\Omega_{\cal
R}|\sim 0.01-0.02$. It must be appreciated that this value for
$\Omega_{\cal R}$ is very large and will still have a significant
dynamical effect. Note that $\Omega_{\cal R} >> \Omega_{\gamma}$,
the energy density of radiation, and $\Omega_{\cal R} \sim
\Omega_{lum}$, the energy density in luminous matter. Indeed, from
nucleosynthesis bounds $\Omega_{\cal R}$ is comparable with
$\Omega_b$, the total contribution of baryons to the normalized
density.

There are two possibilities if $|\Omega_{\cal R}|\sim 0.01-0.02$.
Perhaps cosmological data can be explained without dark energy
through a small backreaction and a reinterpreation of cosmological
observations; for example, many authors have studied observations
in LTB models (albeit with contradictory conclusions), and the
subject is currently under intense investigation \cite{LTBgeo}.
Alternatively, $|\Omega_{\cal R}|\sim 0.01-0.02$, but dark energy
is still needed for consistency with observations. However, a
value of $|\Omega_{\cal R}| \sim 0.01-0.02$ would certainly
necessitate a complete reinvestigation of cosmological
observations. In addition, such a value cannot be naturally
explained by inflation. From standard analysis, depending on the
initial conditions and the details of a specific model of
inflation, $|\Omega_{tot} - 1|$ would be extremely small. Indeed,
any value for $|\Omega_{tot} - 1| >> 10^{-10}$ (say) would be very
difficult to explain within the theory of inflation. Therefore,
inflation indicates that spatial curvature is effectively zero, so
that any non-zero residual curvature (e.g., at the one percent
level) can only be naturally explained in terms of an averaging
effect.

Clearly, averaging can have a very significant dynamical effect on
the evolution of the Universe; the correction terms change the
interpretation of observations so that they need to be accounted
for carefully to determine if a model may be consistent with
observations. Averaging may or may not explain the observed
acceleration. However, it cannot be neglected, and a proper
analysis of cosmological observations will not be possible without
a comprehensive understanding of the affects of averaging. Indeed,
any observation that is based on physics in the (weakly or
strongly) nonlinear regime may well be influenced by the
backreaction effect.

Let us discuss the cosmological observations further.

\begin{itemize}

\item{} The standard analysis of supernova type Ia data and CMB
data in FLRW models cannot be applied directly when backreaction
effects are present, because of the different behaviour of the
spatial curvature \cite{Shapiro:2005}. Indeed, all data needs to
be analysed within a particular inhomogeneous model, and not just
an averaged version. Several studies of the LTB model have
demonstrated that the effect of inhomogeneity on the luminosity
distance can mimic acceleration \cite{LTBgeo}.

\item{}  It is necessary to carefully identify observables
actually measured by an experiment. For example, there are several
different measures of the expansion rate and acceleration. In
addition, there is the question of whether we are dealing with
regional dynamics or global (averaged) dynamics  in any particular
observation. The determination of the Hubble parameter $H_0$ and
the Hubble diagrams from supernovae type Ia (which need data at
low redshift $z<1$) are clearly based on local measurements.
However, most observations of the CMB (e.g., the integrated
Sachs-Wolfe effect, which can be probed at $z>1$) and LSS  (e.g.,
N-body simulations) are sensitive to large-scale (averaged)
properties of the Universe, rather than local ones. Indeed,
integrated affects presumably depend not only on averaged values
of parameters (e.g., ${\cal R}$) but also their actual values over
an appropriate timescale.

\item{} All of our deductions about cosmology from observations
are based on light paths. The presence of inhomogeneities affects
curved null geodesics \cite{NewRas,buch} and can drastically alter
observed distances when they are a sizable fraction of the
curvature radius. The issue of light propagation in such an
inhomogeneous and anisotropic spacetime should be studied in the
context of a realistic quantitative model. In the real Universe,
voids occupy a much larger region as compared to structures
\cite{Hoyle:2003}, hence for local observations light
preferentially travels much more through underdense regions and
the effects of inhomogeneities (voids and structures) on redshift
and luminosity distance are likely to be significant.

\end{itemize}

{\em Acknowledgment:} This work was supported by a grant from the
Natural Sciences and Engineering Research Council of Canada.

\end{document}